%%%%%%%%%%%%%%%%%%%%%%%%%%%%%%%%%%%%%%%%%%%%%%%%%%%%%%%%%%%%%%%%%%%%%%%%%%%%%%
%%%%%%%%%%%%%%%%%%%%%%%%%%%%%%sxlet.tex%%%%%%%%%%%%%%%%%%%%%%%%%%%%%%%%%%%%%%
%%%%%%%%%%%%%%%%%%%%%%%%%%%%%%%%%%%%%%%%%%%%%%%%%%%%%%%%%%%%%%%%%%%%%%%%%%%%%%

%% site dependent options: 
%% \unredoffs and \redoffs define horizontal and vertical offsets 
%% respectively for unreduced and reduced modes. \speclscape defines
%% the \special{} call that sets printer to landscape (sideways) mode.
%% from standard set below, leave uncommented as appropriate or redefine
%
%%% next 400dpi
%\def\unredoffs{} \def\redoffs{\voffset=-.31truein\hoffset=-.48truein}
%\def\speclscape{\special{landscape}}
%
%%% apple lw
%\def\unredoffs{} \def\redoffs{\voffset=-.31truein\hoffset=-.59truein}
%\def\speclscape{\special{ps: landscape}}
%
%%% qms lasergrafix:
%\def\unredoffs{} \def\redoffs{\voffset=-.4truein\hoffset=.125truein}
%\def\speclscape{\special{qms: landscape}}
%
%%% saclay A4 paper:
\def\unredoffs{\hoffset-.14truein\voffset-.2truein} 
 
%\def\speclscape{\special{landscape}}
%
%---------------------------------------------------------------------%
%
\newbox\leftpage \newdimen\fullhsize \newdimen\hstitle \newdimen\hsbody
\tolerance=1000\hfuzz=2pt
\catcode`\@=11 % This allows us to modify PLAIN macros.
%\def\bigans{b }
%\message{ big or little (b/l)? }\read-1 to\answ
%
%\ifx\answ\bigans\message{(This will come out unreduced.}
\magnification=1095\unredoffs\baselineskip=16pt plus 2pt minus 1pt
\hsbody=\hsize \hstitle=\hsize %take default values for unreduced format
%
%\else\message{(This will be reduced.} \let\l@r=L
%\magnification=800\baselineskip=16pt plus 1pt minus 0.5pt \vsize=7truein
%\redoffs \hstitle=8truein\hsbody=4.75truein\fullhsize=10truein\hsize=\hsbody
%
%\output={\ifnum\pageno=0 %%% This is the HUTP version
%  \shipout\vbox{\speclscape{\hsize\fullhsize\makeheadline}
%    \hbox to \fullhsize{\hfill\pagebody\hfill}}\advancepageno}
%  \else
% \almostshipout{\leftline{\vbox{\pagebody\makefootline}}}\advancepageno 
%  \fi}
%\def\almostshipout#1{%\if L\l@r \count1=1 \message{[\the\count0.\the\count1]}
%      \global\setbox\leftpage=#1 \global\let\l@r=R
% \else 
%\count1=2
%  \shipout\vbox{\speclscape{\hsize\fullhsize\makeheadline}
%      \hbox to\fullhsize{\box\leftpage\hfil#1}}  \global\let\l@r=L\fi}
%\fi
%---------------------------------------------------------------------
%
\newcount\yearltd\yearltd=\year\advance\yearltd by -1900

%
% 	restores pagenumbers
%
%       use following instead of \Date on the preliminary draft, 
%       puts date/time on each page in big mode, writes labels in margins

\def\draftmode{\message{ DRAFTMODE }\def\draftdate{{\rm preliminary draft:
\number\month/\number\day/\number\yearltd\ \ \hourmin}}%
\headline={\hfil\draftdate}\writelabels\baselineskip=16pt plus 2pt minus 2pt
 {\count255=\time\divide\count255 by 60 \xdef\hourmin{\number\count255}
  \multiply\count255 by-60\advance\count255 by\time
  \xdef\hourmin{\hourmin:\ifnum\count255<10 0\fi\the\count255}}}
%       use \nolabels to get rid of eqn, ref, and fig labels in draft mode
\def\nolabels{\def\wrlabeL##1{}\def\eqlabeL##1{}\def\reflabeL##1{}}
\def\writelabels{\def\wrlabeL##1{\leavevmode\vadjust{\rlap{\smash%
{\line{{\escapechar=` \hfill\rlap{\sevenrm\hskip.03in\string##1}}}}}}}%
\def\eqlabeL##1{{\escapechar-1\rlap{\sevenrm\hskip.05in\string##1}}}%
\def\reflabeL##1{\noexpand\llap{\noexpand\sevenrm\string\string\string##1}}}
\nolabels
%
% tagged sec numbers
\global\newcount\secno \global\secno=0
\global\newcount\meqno \global\meqno=1
\def\newsec#1{\global\advance\secno by1\message{(\the\secno. #1)}
%\ifx\answ\bigans \vfill\eject \else \bigbreak\bigskip \fi  %if desired
\global\subsecno=0\eqnres@t\noindent{\bf\the\secno. #1}
\writetoca{{\secsym} {#1}}\par\nobreak\medskip\nobreak}
\def\eqnres@t{\xdef\secsym{\the\secno.}\global\meqno=1\bigbreak\bigskip}
\def\sequentialequations{\def\eqnres@t{\bigbreak}}\xdef\secsym{}
\global\newcount\subsecno \global\subsecno=0
\def\subsec#1{\global\advance\subsecno by1\message{(\secsym\the\subsecno. #1)}
\ifnum\lastpenalty>9000\else\bigbreak\fi
\noindent{\bf\secsym\the\subsecno. #1}\writetoca{\string\quad 
{\secsym\the\subsecno.} {#1}}\par\nobreak\medskip\nobreak}
\def\appendix#1#2{\global\meqno=1\global\subsecno=0\xdef\secsym{\hbox{#1.}}
\bigbreak\bigskip\noindent{\bf Appendix #1. #2}\message{(#1. #2)}
\writetoca{Appendix {#1.} {#2}}\par\nobreak\medskip\nobreak}
%
%       \eqn\label{a+b=c}	gives displayed equation, numbered
%				consecutively within sections.
%     \eqnn and \eqna define labels in advance (of eqalign?)
%
\def\eqnn#1{\xdef #1{(\secsym\the\meqno)}\writedef{#1\leftbracket#1}%
\global\advance\meqno by1\wrlabeL#1}
\def\eqna#1{\xdef #1##1{\hbox{$(\secsym\the\meqno##1)$}}
\writedef{#1\numbersign1\leftbracket#1{\numbersign1}}%
\global\advance\meqno by1\wrlabeL{#1$\{\}$}}
\def\eqn#1#2{\xdef #1{(\secsym\the\meqno)}\writedef{#1\leftbracket#1}%
\global\advance\meqno by1$$#2\eqno#1\eqlabeL#1$$}
%
%			 footnotes
\newskip\footskip\footskip14pt plus 1pt minus 1pt %sets footnote baselineskip
\def\footnotefont{\ninepoint}\def\f@t#1{\footnotefont #1\@foot}
\def\f@@t{\baselineskip\footskip\bgroup\footnotefont\aftergroup\@foot\let\next}
\setbox\strutbox=\hbox{\vrule height9.5pt depth4.5pt width0pt}
\global\newcount\ftno \global\ftno=0
\def\foot{\global\advance\ftno by1\footnote{$^{\the\ftno}$}}
%
%say \footend to put footnotes at end
%will cause problems if \ref used inside \foot, instead use \nref before
\newwrite\ftfile   
\def\footend{\def\foot{\global\advance\ftno by1\chardef\wfile=\ftfile
$^{\the\ftno}$\ifnum\ftno=1\immediate\openout\ftfile=foots.tmp\fi%
\immediate\write\ftfile{\noexpand\smallskip%
\noexpand\item{f\the\ftno:\ }\pctsign}\findarg}%
\def\footatend{\vfill\eject\immediate\closeout\ftfile{\parindent=20pt
\centerline{\bf Footnotes}\nobreak\bigskip\input foots.tmp }}}
\def\footatend{}
%
%     \ref\label{text}
% generates a number, assigns it to \label, generates an entry.
% To list the refs on a separate page,  \listrefs
%
\global\newcount\refno \global\refno=1
\newwrite\rfile
\def\ref{[\the\refno]\nref}
\def\nref#1{\xdef#1{[\the\refno]}\writedef{#1\leftbracket#1}%
\ifnum\refno=1\immediate\openout\rfile=refs.tmp\fi
\global\advance\refno by1\chardef\wfile=\rfile\immediate
\write\rfile{\noexpand\item{#1\ }\reflabeL{#1\hskip.31in}\pctsign}\findarg}
%	horrible hack to sidestep tex \write limitation
\def\findarg#1#{\begingroup\obeylines\newlinechar=`\^^M\pass@rg}
{\obeylines\gdef\pass@rg#1{\writ@line\relax #1^^M\hbox{}^^M}%
\gdef\writ@line#1^^M{\expandafter\toks0\expandafter{\striprel@x #1}%
\edef\next{\the\toks0}\ifx\next\em@rk\let\next=\endgroup\else\ifx\next\empty%
\else\immediate\write\wfile{\the\toks0}\fi\let\next=\writ@line\fi\next\relax}}
\def\striprel@x#1{} \def\em@rk{\hbox{}} 
\def\lref{\begingroup\obeylines\lr@f}
\def\lr@f#1#2{\gdef#1{\ref#1{#2}}\endgroup\unskip}
\def\semi{;\hfil\break}
\def\addref#1{\immediate\write\rfile{\noexpand\item{}#1}} %now unnecessary
\def\startrefs#1{\immediate\openout\rfile=refs.tmp\refno=#1}
\def\xref{\expandafter\xr@f}\def\xr@f[#1]{#1}
\def\refs#1{\count255=1[\r@fs #1{\hbox{}}]}
\def\r@fs#1{\ifx\und@fined#1\message{reflabel \string#1 is undefined.}%
\nref#1{need to supply reference \string#1.}\fi%
\vphantom{\hphantom{#1}}\edef\next{#1}\ifx\next\em@rk\def\next{}%
\else\ifx\next#1\ifodd\count255\relax\xref#1\count255=0\fi%
\else#1\count255=1\fi\let\next=\r@fs\fi\next}
%

%
% this is ugly, but moore insists
\newwrite\ffile\global\newcount\figno \global\figno=1
\def\fig{fig.~\the\figno\nfig}
\def\nfig#1{\xdef#1{fig.~\the\figno}%
\writedef{#1\leftbracket fig.\noexpand~\the\figno}%
\ifnum\figno=1\immediate\openout\ffile=figs.tmp\fi\chardef\wfile=\ffile%
\immediate\write\ffile{\noexpand\medskip\noexpand\item{Fig.\ \the\figno. }
\reflabeL{#1\hskip.55in}\pctsign}\global\advance\figno by1\findarg}
\def\xfig{\expandafter\xf@g}\def\xf@g fig.\penalty\@M\ {}
\def\figs#1{figs.~\f@gs #1{\hbox{}}}
\def\f@gs#1{\edef\next{#1}\ifx\next\em@rk\def\next{}\else
\ifx\next#1\xfig #1\else#1\fi\let\next=\f@gs\fi\next}
\newwrite\lfile
{\escapechar-1\xdef\pctsign{\string\%}\xdef\leftbracket{\string\{}
\xdef\rightbracket{\string\}}\xdef\numbersign{\string\#}}

\def\writestop{\def\writestoppt{\immediate\write\lfile{\string\pageno%
\the\pageno\string\startrefs\leftbracket\the\refno\rightbracket%
\string\def\string\secsym\leftbracket\secsym\rightbracket%
\string\secno\the\secno\string\meqno\the\meqno}\immediate\closeout\lfile}}
\def\writestoppt{}\def\writedef#1{}
\def\seclab#1{\xdef #1{\the\secno}\writedef{#1\leftbracket#1}\wrlabeL{#1=#1}}
\def\subseclab#1{\xdef #1{\secsym\the\subsecno}%
\writedef{#1\leftbracket#1}\wrlabeL{#1=#1}}
\newwrite\tfile \def\writetoca#1{}
\def\leaderfill{\leaders\hbox to 1em{\hss.\hss}\hfill}
%	use this to write file with table of contents
\def\writetoc{\immediate\openout\tfile=toc.tmp 
   \def\writetoca##1{{\edef\next{\write\tfile{\noindent ##1 
   \string\leaderfill {\noexpand\number\pageno} \par}}\next}}}
%       and this lists table of contents on second pass

%
\catcode`\@=12 % at signs are no longer letters
%
%	Unpleasantness in calling in abstract and title fonts
\edef\tfontsize{\ifx\answ\bigans scaled\magstep3\else scaled\magstep4\fi}
 \tfontsize  \tfontsize
 \tfontsize \font\titlei=cmmi10 \tfontsize
\font\titleis=cmmi7 \tfontsize \font\titleiss=cmmi5 \tfontsize
\font\titlesy=cmsy10 \tfontsize \font\titlesys=cmsy7 \tfontsize
\font\titlesyss=cmsy5 \tfontsize  \tfontsize
\skewchar\titlei='177 \skewchar\titleis='177 \skewchar\titleiss='177
\skewchar\titlesy='60 \skewchar\titlesys='60 \skewchar\titlesyss='60
 \ifx\answ\bigans\else scaled\magstep1\fi
\ifx\answ\bigans\else

 \font\absi=cmmi10 scaled\magstep1
\font\absis=cmmi7 scaled\magstep1 \font\absiss=cmmi5 scaled\magstep1
\font\abssy=cmsy10 scaled\magstep1 \font\abssys=cmsy7 scaled\magstep1
\font\abssyss=cmsy5 scaled\magstep1 
\skewchar\absi='177 \skewchar\absis='177 \skewchar\absiss='177
\skewchar\abssy='60 \skewchar\abssys='60 \skewchar\abssyss='60
\fi
\font\ninerm=cmr9 \font\sixrm=cmr6 \font\ninei=cmmi9 \font\sixi=cmmi6 
\font\ninesy=cmsy9 \font\sixsy=cmsy6 \font\ninebf=cmbx9 
\font\nineit=cmti9 \font\ninesl=cmsl9 \skewchar\ninei='177
\skewchar\sixi='177 \skewchar\ninesy='60 \skewchar\sixsy='60 
\def\ninepoint{\def\rm{\fam0\ninerm}% switch to footnote font
\textfont0=\ninerm \scriptfont0=\sixrm \scriptscriptfont0=\fiverm
\textfont1=\ninei \scriptfont1=\sixi \scriptscriptfont1=\fivei
\textfont2=\ninesy \scriptfont2=\sixsy \scriptscriptfont2=\fivesy
\textfont\itfam=\ninei \def\it{\fam\itfam\nineit}\def\sl{\fam\slfam\ninesl}%
\textfont\bffam=\ninebf \def\bf{\fam\bffam\ninebf}\rm} 
%
%---------------------------------------------------------------------
%

\hyphenation{anom-aly anom-alies coun-ter-term coun-ter-terms}
\def\inv{^{\raise.15ex\hbox{${\scriptscriptstyle -}$}\kern-.05em 1}}

\def\Dsl{\,\raise.15ex\hbox{/}\mkern-13.5mu D} %this one can be subscripted
\def\dsl{\raise.15ex\hbox{/}\kern-.57em\partial}

 %pound sterling
\def\lspace{\ifx\answ\bigans{}\else\qquad\fi}
\def\lbspace{\ifx\answ\bigans{}\else\hskip-.2in\fi} % $$\lbspace...$$
\def\boxeqn#1{\vcenter{\vbox{\hrule\hbox{\vrule\kern3pt\vbox{\kern3pt
	\hbox{${\displaystyle #1}$}\kern3pt}\kern3pt\vrule}\hrule}}}
\def\mbox#1#2{\vcenter{\hrule \hbox{\vrule height#2in
		\kern#1in \vrule} \hrule}}  %e.g. \mbox{.1}{.1}
%	matters of taste
%\def\tilde{\widetilde} \def\bar{\overline} \def\hat{\widehat}
%
% some sample definitions
  %     curly letters

\def\darr#1{\raise1.5ex\hbox{$\leftrightarrow$}\mkern-16.5mu #1}
 %pound sterling

 %puts a small half in a displayed eqn
 %ditto 4/9
 %ditto 3/2
\def\roughly#1{\raise.3ex\hbox{$#1$\kern-.75em\lower1ex\hbox{$\sim$}}}

\def\p2inf{\mathrel{\mathop{\sim}\limits_{\scriptscriptstyle
{p^2 \rightarrow \infty }}}}
\def\kap2inf{\mathrel{\mathop{\sim}\limits_{\scriptscriptstyle
{\kappa \rightarrow \infty }}}}
\def\x2inf{\mathrel{\mathop{\sim}\limits_{\scriptscriptstyle
{x \rightarrow \infty }}}}
\def\Lam2inf{\mathrel{\mathop{\sim}\limits_{\scriptscriptstyle
{\Lambda \rightarrow \infty }}}}
\def\frac#1#2{{{#1}\over {#2}}}

\def\Gev{{\rm GeV}}

\def\lsim{\mathrel{mathpalette\@v1000ersim<}}
\def\gsim{\mathrel{mathpalette\@versim>}}

\catcode`@=11 %This allows us to modify plain macros
\def\slash#1{\mathord{\mathpalette\c@ncel#1}}
 \def\c@ncel#1#2{\ooalign{$\hfil#1\mkern1mu/\hfil$\crcr$#1#2$}}
\def\lsim{\mathrel{\mathpalette\@versim<}}
\def\gsim{\mathrel{\mathpalette\@versim>}}
 \def\@versim#1#2{\lower0.2ex\vbox{\baselineskip\z@skip\lineskip\z@skip
       \lineskiplimit\z@\ialign{$\m@th#1\hfil##$\crcr#2\crcr\sim\crcr}}}
\catcode`@=12 %at signs are no longer letters

\def\PR{{\it Phys.~Rev.~}}

\def\NP{{\it Nucl.~Phys.~}}
\def\PL{{\it Phys.~Lett.~}}

\def\SJNP{{\it Sov.~Jour.~Nucl.~Phys.~}}
\def\ZP{{\it Zeit.~Phys.~}}

\def\vol#1{{\bf #1}}
\def\vyp#1#2#3{\vol{#1} (#2) #3}

\def\Asl{\raise.15ex\hbox{/}\mkern-11.5mu A}
\def\psl{\lower.12ex\hbox{/}\mkern-9.5mu p}
\def\qsl{\lower.12ex\hbox{/}\mkern-9.5mu q}
\def\rsl{\lower.03ex\hbox{/}\mkern-9.5mu r}
\def\ksl{\raise.06ex\hbox{/}\mkern-9.5mu k}

%%%%%%%%%%%%%%%%%%%%%%%%%%%%%%%%%%%%%%%%%%%%%%%%%%%%%%%%%%%%%%%%%%%%%%%%%%%%%%%

\pageno=0\nopagenumbers\tolerance=10000\hfuzz=5pt
\line{\hfill RAL-96-088}
\vskip 36pt
\centerline{\bf Renormalization Scheme Consistent Structure Functions,}
\vskip 6pt
\centerline{\bf Including Leading $\ln (1/x)$ Terms.}
\vskip 36pt
\centerline{Robert~S.~Thorne}
\vskip 12pt
\centerline{\it Rutherford Appleton Laboratory,}
\centerline{\it Chilton, Didcot, Oxon., OX11 0QX, U.K.}
\vskip 0.9in
{\narrower\baselineskip 10pt
\centerline{\bf Abstract}
\medskip
We present calculations of structure functions
using a renormalization scheme consistent
expansion which is leading order in both $\ln(1/x)$ and $\alpha_s(Q^2)$.
There is no factorization scheme dependence, and the  
``physical anomalous dimensions'' of Catani naturally appear. 
A relationship between the small $x$ 
forms of the inputs $F_2(x,Q_0^2)$ and $F_L(x,Q_0^2)$ is predicted.
Analysis of a very wide range of data for $F_2(x,Q^2)$
is performed, and a very good global fit obtained. 
The prediction for $F_L(x,Q^2)$ produced using this method 
is smaller than the usual 
NLO in $\alpha_s(Q^2)$ predictions for $F_L(x,Q^2)$, and different in shape.}
   
\vskip 0.7in
\line{RAL-96-088\hfill}
\line{October 1996\hfill}
\vfill\eject
\footline={\hss\tenrm\folio\hss}

%%%%%%%%%%%%%%%%%%%%%%%%%%%%%%%%%%%%%%%%%%%%%%%%%%%%%%%%%%%%%%%%%%%%%%%%%%%%%%%

%%%%%%%%%%%%%%%%%%%%%%%%%%%%%%%%%%%%%%%%%%%%%%%%%%%%%%%%%%%%%%%%%%%%%%%%%%%%%%%

\newsec{Introduction}

The recent measurements of $F_2(x,Q^2)$ at HERA have provided data on
a structure function at far lower values of $x$ than any previous
experiments, and show that there is a marked rise in $F_2(x,Q^2)$ at
very small $x$ down to rather low values of $Q^2$ \ref\hone{H1 collaboration,
\NP B \vyp{470}{1996}{3}.}\ref\zeus{ZEUS collaboration: M. Derrick {\it et al},
\ZP C \vyp{69}{1996}{607}\semi Preprint DESY 96-076 (1996), to be published
in \ZP C.}. Indeed, the most recent
measurements demonstrate that the rise persists for values of
$Q^2$ as low as 1.5 $\Gev^2$. These measurements have led to a great deal of 
interest in how one should best calculate structure functions.

The particular interest in the small $x$ region comes about 
because it has long been known that there is potential small $x$ 
enhancement of the structure functions at high orders in the strong coupling
constant \ref\BFKL{L.N. Lipatov, \SJNP \vyp{23}{1976}{338}\semi
E.A. Kuraev, L.N. Lipatov and V.S. Fadin, {\it Sov.~Jour. JETP} 
\vyp{45}{1977}{199}\semi
Ya. Balitskii and L.N. Lipatov, \SJNP \vyp{28}{1978}{6}.} which is known as 
BFKL physics: i.e. the effective 
splitting function governing the growth of the gluon Green's function 
at small $x$ is of the form $P(x)=\sum_{m=1}^{\infty} a_m x^{-1}
\alpha_s^m\ln^{m-1}(1/x)$ \ref\Jar{T. Jaroszewicz, \PL B 
\vyp{116}{1982}{291}.}, where the $a_m$ are such that an asymptotic 
growth $x^{-1-\bar\alpha_s 4\ln2}$ was predicted ($\bar \alpha_s
=3 \alpha_s/\pi$). This implies that one  
needs more than the normal fixed order in $\alpha_s$ expansion in order to
describe physics at small $x$. Qualitative studies incorporating these ideas 
to obtain the structure functions (rather than just the gluon Green's 
function) were in reasonable qualitative agreement 
with early HERA data \ref\KMS{A.J. Askew, J. Kwiecinski, 
A.D. Martin and P.J. Sutton, \PR D \vyp{49}{1994}{4402}\semi
A.J. Askew {\it et al}, \PL B \vyp{325}{1994}{212}}.  

However, it was shown 
to be possible to obtain very good fits to the same 
data whilst ignoring these leading $\ln (1/x)$ terms.
Using the Altarelli--Parisi 
evolution equations at next to leading (NLO) order,
or even leading order (LO) in $\alpha_s$  
and starting with flat \ref\DAS{R.D. Ball and S. Forte, \PL B 
\vyp{335}{1994}{77}; \PL B 
\vyp{336}{1994}{77}.}, or even valence--like inputs \ref\GRV{M. Gl\"uck, 
E. Reya and A. Vogt, \ZP C \vyp{48}{1990}{471}; \ZP C \vyp{53}{1992}{127};
\PL B \vyp{306}{1993}{391}; \ZP C \vyp{67}{1995}{433}.},
predicted a steep (though not powerlike) 
rise in the small $x$ structure functions at scales reasonably 
far above the starting scale $Q_0^2$, and gave a good fit
as long as $Q_0^2$ was chosen 
to be low.  This countered the BFKL approach 
which, after all, was derived using
a less well--defined theoretical framework than the renormalization
group approach based on the factorization of collinear
singularities and ignored everything but the leading $\ln (1/x)$ terms. 
An approach somewhat intermediate between these extremes is also used,
i.e. the fixed order evolution beginning
from inputs for the parton distributions of the form $x^{-1-\lambda}$ at small
$x$, with $\lambda \sim 0.2$, partially justifying 
the relatively steep input (significantly steeper than that expected from 
non-perturbative physics, if somewhat less than $\bar\alpha_s 4\ln2$) 
from BFKL physics, e.g. \ref\MRSi{A.D. Martin, W.J. Stirling and R.G. Roberts, 
\PL B \vyp{354}{1995}{155}\semi H.L. Lai {\it et al},
\PR D \vyp{51}{1995}{4763}.}. 

Recently, due to the work of 
Catani and Hautmann \ref\CatHaut{S. Catani and F. Hautmann, \PL B 
\vyp{315}{1993}{157}; \NP B \vyp{427}{1994}{475}.}, it is possible to 
use the leading in $\ln (1/x)$ 
expressions for the anomalous dimensions and coefficient functions within 
the well--defined renormalization group approach. 
Using the $k_T$-factorization theorem \ref\kti{S. Catani, M. Ciafaloni 
and F. Hautmann, \PL B \vyp{242}{1990}{97}; \NP B \vyp{366}{1991}{135};
\PL B \vyp{307}{1993}{147}.}\ref\ktii{J.C. Collins and R.K. Ellis, \NP B
\vyp{360}{1991}{3}.}\CatHaut\ they verified the form of the gluon 
anomalous dimensions $\gamma^0_{gg}(N,\alpha_s)$ and 
$\gamma^0_{fg}(N,\alpha_s)$
(where $\gamma(N)=\int_{0}^{1}x^N P(x) d\,x$) and derived expressions for
$\gamma^1_{ff}(N,\alpha_s(Q^2))$ and $\gamma^1_{fg}(N,\alpha_s(Q^2))$
in certain factorization schemes 
(since $\gamma^0_{ff(g)}(N,\alpha_s(Q^2))=0$ the quark anomalous dimensions 
are a power of $N^{-1}$ down on those of the gluon). They also derived 
expressions for
the coefficient functions $C^g_{L,1}(N,Q^2)$, $C^f_{L,1}(N,Q^2)$, 
$C^g_{2,1}(N,Q^2)$ and $C^f_{2,1}(N,Q^2)$ (all
zeroth order quantities being zero except $C^f_{2,0}$, which is
unity). This facilitated calculations of structure functions within
the normal renormalization group framework whilst including the leading 
$\ln(1/x)$ terms. Calculations were performed using these 
terms \ref\fsdepbf{R.D. Ball and S. Forte, \PL B \vyp{351}{1995}{313}; \PL B
\vyp{358}{1995}{365}.}\nref\fsdepehw{R.K. Ellis, F. Hautmann and B.R. Webber,
\PL B \vyp{348}{1995}{582}.}--\ref\fsdepfrt{J.R. Forshaw, R.G. Roberts and 
R.S. Thorne, \PL B \vyp{356}{1995}{79}.} and comparisons
with data made. The calculations used different methods of solution,
made rather different assumptions and used (different) ans\"atze for unknown
terms. Consequently, differing results were obtained. The conclusions
which could be drawn regarding the inclusion of the leading $\ln
(1/x)$ terms depended very much on which approach was
taken. However, including these terms did not seem to
improve the best fits for the small $x$ data using one-- or
two--loop evolution \fsdepbf\fsdepfrt. Indeed, many ways of inclusion 
significantly worsened the fits, particularly if
they were more global, i.e. constrained by large $x$ data \fsdepbf.
Also, there seemed to be a very strong
dependence on the factorization scheme used in the
calculations when including the leading $\ln (1/x)$ terms \fsdepbf\fsdepfrt
\ref\sd{S. Catani, \ZP C \vyp{70}{1996}{263}.}. 

The high precision of the most recent HERA data constrains theory far more 
than previously. The best recent global fits seem to come from those 
intermediate approaches which 
use NLO perturbation theory with a quite steep (unexplained) 
input for the singlet quark with 
$\lambda \sim 0.2$ and a similar form of small $x$ input for the 
gluon \ref\MRSii{A.D. Martin, W.J. Stirling and R.G. Roberts, 
preprint DTP/96/44 or RAL-TR-96-037, May 1996,
to be published in \PL B.} (unless $Q_0^2$ is less than  
$\sim 4\Gev^2$, in which case it must be flatter or even valence--like).
Fixed order perturbation theory with 
flat or valence--like inputs and low $Q_0^2$ fails
at the lowest $x$ values, and for fits to the small $x$ data alone relatively
steep inputs for
the singlet quark, i.e. $\lambda \gsim 0.2$, seem to be required 
\ref\steep{R.D. Ball and S.Forte, preprint Edinburgh 96/10
or DFTT 36/96, June 1996, to appear in proceedings of DIS 96, 
(Rome, April 1996).}.
Approaches including the resummed terms now seem to 
fail \ref\fail{S. Forte and R.D. Ball, preprint Edinburgh 96/14
or DFTT 35/96, July 1996, to appear in proceedings of DIS 96, 
(Rome, April 1996)\semi I. Bojak and M. Ernst, DO-TH 96/18, September 1996.}
in practically all factorization schemes.  

In this paper we will demonstrate that the apparent failure of approaches
using the leading $\ln (1/x)$ terms, and certainly the factorization scheme
dependence, is due to incorrect methods of incorporating these terms.
The correct leading order renormalization scheme consistent (RSC) calculation 
naturally includes leading $\ln (1/x)$ terms in a form which has already been
derived by Catani and called ``physical anomalous 
dimensions'' \ref\Cat{S. Catani,
talk at UK workshop on HERA physics, September 1995, unpublished;
preprint DDF 248/4/96, April 1996; preprint DDF 254-7-96, July 1996,
to appear in proceedings of DIS 96, (Rome, April 1996).}. It also 
provides some limited predictive power at small $x$. We will 
discuss this method of calculation, then make detailed
comparisons to data, and with the aid of the new HERA data demonstrate 
that this calculation leads to a very good global fit to all 
$F_2(x,Q^2)$ data. Indeed, the complete RSC calculation, including leading
$\ln (1/x)$ terms, is clearly preferred by the latest data.

\newsec{The Renormalization Scheme Consistent Expansion.}

For simplicity we work in moment space in this paper, i.e. 
define the moment space structure functions by the Mellin transformation, 
\eqn\melltranssf{F(N,Q^2)= \int_0^1\,x^{N-1}{\cal F}(x,Q^2)dx.}
and similarly for the coefficient function. As
with the anomalous dimension we define the moment
space parton distribution as the Mellin transformation
of a rescaled parton density i.e 
\eqn\melltranspd{f(N,Q^2)= \int_0^1\,x^{N}{\rm f}(x,Q^2)dx.}
The moment space expression for a structure function is then
\eqn\strcfunc{F(N,Q^2)= \sum_a
C^a(N,\alpha_s(Q^2))f_a(N,Q^2),}
and the parton distributions evolve according to the 
perturbative renormalization group equation
\eqn\apeqn{{d \, f_a(Q^2) \over d \ln Q^2}= \sum_b
\gamma_{ab}(\alpha_s(Q^2)) f_b(Q^2),}
where we choose both the factorization and renormalization scale 
to be equal to $Q^2$.
The coefficient functions and anomalous dimensions are factorization 
scheme dependent, of course. 

There are two independent structure functions 
$F_2(N,Q^2)$ and $F_L(N,Q^2)$ which have both singlet
and nonsinglet contributions. In general we may write
\eqn\srucdef{F_i(N,Q^2) =
{1\over N_f}\biggl(\sum_{j=1}^{N_f}e^2_j\biggr)F^S_i(N,Q^2) +
F^{NS}_{i}(N,Q^2) \hskip 0.5in (i=2,L),}
where in terms of coefficient functions and parton densities 
\eqn\fsing{\eqalign{F^S_i(N,Q^2) &= C^f_i(N,\alpha_s)f^S(N,Q^2) +
C^g_i(N,\alpha_s)g(N,Q^2),\cr
F^{NS}_i(N,Q^2) &= C^{NS}_i(N,\alpha_s)\sum_{j=1}^{N_f}
e_j^2 f^{NS}_{q_j}(N,Q^2).\cr}}
$N_f$ is the number of active quark flavours (we only consider 
massless quarks), and $f^S(N,Q^2)$ and
$f^{NS}_{q_j}(N,Q^2)$ are the singlet and nonsinglet quark
distribution functions respectively.

\medskip

In order to devise an expansion scheme for the calculation of these 
structure functions which is useful at both large
and small $x$, we would {\it a priori} expect that we would need to use
the known anomalous dimensions and coefficient
functions at low orders in both $\alpha_s$ and in the leading $\ln (1/x)$
expansion. There have already been some methods along these lines; however, 
these have suffered from 
scheme dependence. This is clearly incorrect since we do not expect 
factorization scheme dependence in a well ordered calculation of a physical 
quantity. 

As already mentioned, Catani has shown how one 
may obtain factorization scheme independent results, even at small $x$, by the 
use of factorization scheme independent, or physical, 
anomalous dimensions. We refer to his papers \Cat, or for a slightly
different presentation \ref\rst{R.S. Thorne, preprint RAL-TR-96-065, 
to appear.}, for details. Very briefly, using \strcfunc\ one writes 
parton distributions in terms of structure functions and coefficient 
functions, and substituting into \apeqn\ leads to evolution equations for the 
structure functions themselves in terms of physical anomalous dimensions
$\Gamma_{22}(N,\alpha_s)$, $\Gamma_{2L}(N,\alpha_s)$, etc. These can be 
expressed in terms of coefficient functions and anomalous dimensions in  
any factorization scheme, but in scheme independent combinations.
In order to perform our calculation of structure functions we do not 
work with these physical anomalous dimensions from the outset. Rather we 
will simply demand a complete leading order, 
renormalization scheme consistent (RSC) 
calculation. This leads to expressions which are
unique, up to nonperturbative inputs, and which 
naturally contain the physical anomalous dimensions. 

\medskip

To begin, let us consider what we normally mean by 
``consistency with renormalization scheme dependence''.
In the loop expansion the order
of a term is determined simply by its order in $\alpha_s$, and in the 
leading $\ln(1/x)$ expansion $(\ln(1/x))^{-1}$ is put on an equal footing 
to $\alpha_s$. In both forms of
expansion one demands that, once we choose a particular renormalization
scheme and work to a particular order in this renormalization
scheme, we include all terms in our expressions for the 
structure function which are of lower order than the uncertainty
due to the freedom of choice of renormalization scheme (i.e. the uncertainty 
in the definition of the coupling constant), and no others.\foot {Following 
this 
prescription one automatically obtains factorization scheme independent 
expressions in both the above 
expansion schemes. This is well known in the loop 
expansion, see \rst\ for a discussion of this and the more complicated case
of the leading $\ln (1/x)$ expansion.} If working with the $n$-loop coupling
constant, the ambiguity in its defintion due to renormalization scheme
uncertainty is of order $\alpha_s^{n+1}$. Thus, the uncertainty when 
working to $n$-th order is the change in the 
leading order expression under the change of coupling 
$\alpha_s \to \alpha_s(1+\epsilon\alpha_s^n)$. 
Hence, the uncertainty in the whole structure function
is of the order of the change of the leading
order part under such a change in the coupling, and the $n$-th order 
renormalization scheme independent expression includes all complete terms
of lower order than this change. 

This definition gives a well--defined
way of building up a solution to the structure functions, 
but relies upon the definition of a given expansion scheme.
It leaves an ambiguity about how we define leading order 
expressions and about how we define the order of terms compared to this 
leading order. The loop expansion and leading $\ln(1/x)$ 
expansion are just the two most commonly used examples. 
Both have potential problems: in the former one does not 
worry about the fact that the large $\ln(1/x)$ terms can cause enhancement 
of terms which are higher order in $\alpha_s$ at small $x$, 
and in the latter one does not worry about the fact that 
at large $x$, especially as $Q^2$ increases, it is the terms that are of 
lowest order in $\alpha_s$ which are most important. Hence, one would 
think that both have limited regions of validity. 

The shortcomings of these two expansion methods come about because even  
though a given order contains no terms 
which are inconsistent with working to this order in a particular
renormalization scheme, in neither does it include every one of the terms 
which are consistent with working to a given order in the
renormalization scheme. In each expansion scheme some of the terms 
appearing at what we call higher orders are not actually
subleading in $\alpha_s$ to any terms which have already appeared. Thus,
although (for a given expansion method) these terms are 
formally of the same order as 
uncertainties due to the choice of renormalization scheme, they are not terms 
which are actually generated by changes in renormalization 
scheme.\foot{Similarly, they are not generated by changes in 
renormalization scale.} 

In order to demonstrate this point more clearly we consider a simple toy 
model. Let us imagine some hypothetical physical quantity which can be
expressed in the form
\eqn\hypothet{H(N,\alpha_s(Q^2))=\sum_{m=1}^{\infty}\alpha_s(Q^2)
\sum_{n=-m}^{\infty}a_{mn}N^n\equiv \sum_{i=0}^{\infty}\alpha^i_s(Q^2)
\sum_{j=1-i}^{\infty}b_{ij}\biggl({\alpha_s(Q^2)\over N}\biggr)^j.}
The first way of writing $H(N,\alpha_s(Q^2))$ as a power series corresponds to
the loop expansion, where we work order by order in $m$, out to $m=k$,  
and use the $k$-loop
coupling. The second corresponds to the leading $\ln(1/x)$ expansion, 
where we work order by order in $i$, out to $i=l$, and use the $(l+1)$-loop 
coupling.
Let us, for a moment, 
consider the leading order expression in the loop expansion,
$\alpha_s(Q^2) \sum_{n=-1}^{\infty}a_{1n}N^n$.
The coupling is uncertain by ${\cal O}(\alpha^2_s(Q^2))$ and hence the 
uncertainty of the leading order expression (i.e. the change due to 
a change of the coupling) is $\sim\alpha_s^2(Q^2)\sum_{n=-1}^{\infty}
b_{1n}N^n$. We see that there is no change of any sort with a 
power of $N$ less than $-1$, and hence any such term is not really subleading.
Similarly, the uncertainty of the leading order expression in the leading
$\ln (1/x)$ expansion contains no terms at first order in $\alpha_s$ (or
with positive powers of $N$), and such terms are not really 
subleading either. The full set of terms in the combination
of both leading order expressions is genuinely leading order, and 
renormalization scheme independent by definition. 

Perhaps the best way in which to write our expression for 
$H(N,\alpha_s(Q^2))$ in order to appreciate these points is 
\eqn\hypotheti{H(N,\alpha_s(Q^2))=\sum_{m=-1}^{\infty}N^m
\sum_{n=1}^{\infty}c_{mn}\alpha_s^n(Q^2)+ \sum_{m=2}^{\infty}N^{-m}
\sum_{n=m}^{\infty}c_{mn}\alpha_s^n(Q^2),}
i.e. as an infinite number of power series in $\alpha_s(Q^2)$, one 
for each power of $N$. Each of these series in $\alpha_s(Q^2)$ is 
independent of the others, and the lowest order in $\alpha_s(Q^2)$ of each 
is therefore renormalization scheme independent and part of
the complete leading order expression for $H(N,\alpha_s(Q^2))$.  
The full leading order 
expression for $H(N,\alpha_s(Q^2))$ is therefore
\eqn\hypothetlo{\eqalign{H_0(N,\alpha_s(Q^2))&=\sum_{m=-1}^{\infty}N^m
c_{m1}\alpha_s(Q^2)+ \sum_{m=2}^{\infty}c_{mm}N^{-m}
\alpha^{m}_s(Q^2)\cr
&\equiv \alpha_s(Q^2)
\sum_{n=-1}^{\infty}a_{0n}N^n+ 
\sum_{j=2}^{\infty}b_{0j}\biggl({\alpha_s(Q^2)\over N}\biggr)^j.}}
Hence, the combined set of terms considered to be leading order in both 
the previous expansion schemes comprise the full set of renormalization scheme
invariant, and thus truely leading order, terms.
By considering $H(N,\alpha_s(Q^2))$ written as \hypotheti, 
and considering a redefinition of the coupling constant,
$\alpha_s(Q^2)\to \alpha_s(Q^2)+{\cal O}(\alpha^m_s(Q^2))$, we see that the 
$n$-th order expression for $H(N,\alpha_s(Q^2))$, which should be used with 
the $n$-loop coupling constant, is the sum of the first $n$ terms in
each of the power series in $\alpha_s(Q^2)$.

Similar arguments have already been applied to the anomalous dimensions and 
coefficient functions (\kti\ and particularly \fsdepbf).
Here we take a strong viewpoint and
insist that the complete renormalization scheme consistent expressions,
with no artificial supression of leading $\ln(1/x)$ terms,
must be used. Futhermore, the expressions must be those
for the physical structure functions, not for the factorization 
and renormalization scheme dependent coefficient functions and anomalous 
dimensions.

\medskip

When considering the real structure functions the situation is 
a great deal more complicated technically than the toy model, 
but the principle is the same. 
One complication is that the structure functions 
are combinations of perturbative evolution parts and input parts 
(which are perturbative with nonperturbative 
factors), rather than one simple power series in $\alpha_s(Q^2)$. However,
the physical 
consequence of the factorization theorem is that there is some fixed 
nonperturbative factor for the structure functions which we cannot 
calculate in perturbation theory, but we can 
predict the perturbative evolution of the structure functions in terms of 
this factor. Thus, we choose two independent variables for each 
structure function, the input at some starting scale $Q_0^2$, and the evolution
away from this starting scale. We calculate the lowest order 
RSC expression for each of these, and combine to give the 
full LO expression. In this paper we only do this for the singlet 
structure function, since this is hugely dominant at small $x$. 
The procedure is also complicated here because the evolution 
of $F^S_2$ and $F^S_L$ is coupled, but the 
expression for each is a sum of terms consisting of input and evolution parts, 
and for each term we take the LO expression for the input and for 
the evolution. 

Using the above prescription it is relatively straightforward, but rather 
involved to calculate the full leading order  
RSC expressions. The full details appear in \rst, here we just present 
the results. In order to do this most succinctly we express the result 
in terms of physical quantities. Hence, in order to explain notation 
(which is similar to that in \Cat),
and also slightly elucidate the form of the final expressions we 
consider the solution to the 1-loop renormalization group equation first
(this being part of the full solutions).
Using boundary conditions $\hat F^{0,l}_L(N,Q_0^2)=\hat F_L(N)$ and
$F^{0,l}_2(N,Q_0^2)=F_2(N)$, where $F_i(N)$ are nonperturbative inputs,
and the superscript $0,l$ denotes one--loop quantities,
we may write the solution for the longitudinal
structure function as
\eqn\fullsolii{\hat F^{0,l}_L(N,Q^2)=\hat F^{0,l,+}_L(N)\biggl({\alpha_s(Q_0^2)
\over \alpha_s(Q^2)}\biggr)^{\tilde \Gamma^{0,l,+}(N)}+\hat F^{0,l,-}_L(N)
\biggl({\alpha_s(Q_0^2)
\over \alpha_s(Q^2)}\biggr)^{\tilde \Gamma^{0,l,-}(N)},}
where $\hat F_L(N,Q^2)=F_L(N,Q^2)/(\alpha_s(Q^2)/2\pi)$, 
$\tilde \Gamma^{0,l,+(-)}(N)$ are the two eigenvalues of the 
${\cal O}(\alpha_s)$ 
physical anomalous dimension matrix, which are the same as those
of the ${\cal O}(\alpha_s)$ parton anomalous dimension matrix, divided by
$b_0 \alpha_s(Q^2)$. (The superscript $S$ is dropped for the rest of 
this section.) Having chosen to write the  
solution for $F^{0,l}_L(N,Q^2)$ in this way we may then 
write $F^{0,l}_2(N,Q^2)$ as
\eqn\fullsoliii{F^{0,l}_2(N,Q^2)=e^+(N)\hat F^{0,l,+}_L(N)
\biggl({\alpha_s(Q_0^2)
\over \alpha_s(Q^2)}\biggr)^{\tilde\Gamma^{0,l,+}(N)}+e^-(N)\hat F^{0,l,-}_L(N)
\biggl({\alpha_s(Q_0^2)
\over \alpha_s(Q^2)}\biggr)^{\tilde \Gamma^{0,l,-}(N),}}
where $e^+(N)\hat F^{0,l,+}_L(N)+e^-(N)\hat F^{0,l,-}_L(N)=F_2(N)$.
The $e^{+(-)}(N)$ come from the eigenvectors of the ${\cal O}(\alpha_s)$
parton anomalous dimension and the ${\cal O}(\alpha_s)$ longitudinal 
coefficient functions. In practice
\eqn\fullsolc{\eqalign{\hat F^{0,l,+}_L(N)&=\hat F_L(N)-
{36-8N_f\over 27}F_2(N)+{\cal O}(N),\hskip 0.2in \hat F^{0,l,-}_L(N)={36-8N_f
\over 27} F_2(N)+{\cal O}(N),\cr 
F^{0,l,+}_2(N)
&={N\over 6}\biggl(\hat F_L(N)-\biggl({36-8N_f\over 27}\biggr)\biggr)F_2(N)
+{\cal O}(N^2),\hskip 0.2in  
F^{0,l,-}_2(N)=F_2(N)+{\cal O}(N),}} 
where $F_2^{0,l,+(-)}(N)=e^{+(-)}(N)\hat F_L^{0,l,+(-)}(N)$.

Taking into account the leading $\ln (1/x)$ terms as well, the 
expressions acquire additional terms. Explicitly we obtain 
\eqn\fullsolfl{\eqalign{F_{L,RSC,0}(N&,Q^2)
={\alpha_s(Q_0^2)\over 2\pi}\Biggl[\biggl({\alpha_s(Q_0^2)\over
\alpha_s(Q^2)}\biggr)^{\tilde \Gamma^{0,l,+}(N)-1}
\exp\biggl[\int_{\alpha_s(Q^2)}^{\alpha_s(Q_0^2)}
{\Gamma^{\tilde 0}_{LL}(N,\alpha_s(q^2))\over b_0\alpha^2_s(q^2)}
d\alpha_s(q^2)\biggr]\times\cr
&\biggl(\hat F^{0,l,+}_L(N)+\biggl(\hat F_L(N)-\biggl({36-8N_f\over 27}\biggr)
F_2(N)\biggr)(\exp[\ln(Q_0^2/A_{LL})
\Gamma^0_{LL}(N,\alpha_s(Q_0^2))]-1)\biggr)\cr
&+\hat F^{0,l,-}_L(N)\biggl({\alpha_s(Q_0^2)\over
\alpha_s(Q^2)}\biggr)^{\tilde \Gamma^{0,l,-}(N)-1}\Biggr],}}
for the longitudinal structure function, and
\eqn\fullsolfderiv{\eqalign{\biggl({d\,F_{2}(N,Q^2)\over d\,\ln Q^2}
\biggr)_{RSC,0}&
=\alpha_s(Q_0^2)\Biggl[e^-(N)\Gamma^{0,l,-}(N)
\hat F^{0,l,-}_L(N)\biggl({\alpha_s(Q_0^2)\over
\alpha_s(Q^2)}\biggr)^{\tilde \Gamma^{0,l,-}(N)-1}\cr
&+\biggl(e^+(N)\Gamma^{0,l,+}(N)\hat F^{0,l,+}_L(N)- \Gamma^{1,0}_{2,L}(N)
\biggl(\hat F_L(N)-\biggl({36-8N_f\over 27}\biggr)F_2(N)\biggr)\cr
&\hskip-0.6in+\Gamma^{1}_{2L}(N,\alpha_s(Q_0^2))
\biggl(\hat F_L(N)-\biggl({36-8N_f\over 27}\biggr)F_2(N)\biggr)
\exp[\ln(Q_0^2/A_{LL})
\Gamma^0_{LL}(N,\alpha_s(Q_0^2))]\biggr)\times\cr
&\exp\biggl[
\int_{\alpha_s(Q^2)}^{\alpha_s(Q_0^2)}
{\Gamma^{\tilde 0}_{LL}(N,\alpha_s(q^2))\over b_0\alpha^2_s(q^2)}
d\alpha_s(q^2)\biggr]\biggl({\alpha_s(Q_0^2)\over
\alpha_s(Q^2)}\biggr)^{\tilde \Gamma^{0,l,+}(N)-1}\Biggr],\cr}}
and
\eqn\fullsolfin{\eqalign{F_{2,RSC,0}(N,Q_0^2) &=F_2(N)\cr 
&\hskip -0.8in + \alpha_s(Q_0^2)
{\Gamma^1_{2L}(N,\alpha_s(Q_0^2))\over\Gamma^0_{LL}(N,\alpha_s(Q_0^2))}
\biggl(\hat F_L(N)-{(36-8N_f)\over 27}
F_2(N)\biggr)(\exp[\ln(Q_0^2/A_{LL})\Gamma^0_{LL}(N,\alpha_s(Q_0^2))]-1)\cr
&\hskip -0.8in+\ln(Q_0^2/A_{LL})
\alpha_s(Q_0^2)\biggl(e^+(N)\Gamma^{0,l,+}(N)\hat 
F_L^{0,l,+}(N,Q_0^2) + e^-(N)\Gamma^{0,l,-}(N)\hat F_L^{0,l,-}(N,Q_0^2) \cr
&\hskip 1.2in -\Gamma^{1,0}_{2,L}(N)\biggl(
\hat F_L(N)-\biggl({36-8N_f\over27}\biggr)F_2(N)\biggr)\biggr).}}
for $F_2(N,Q^2)$. 

We should explain the terms in these expressions. 
$\Gamma^{0}_{LL}(N,\alpha_s)$ is the gluon anomalous dimension at 
leading order in $\ln (1/x)$, which it turns out governs the small 
$x$ evolution 
of $F_L(N,Q^2)$, as seen in \fullsolfl. It is also identical to the physical
anomalous dimension, hence the notation. 
$\Gamma^{\tilde 0}_{LL}(N,\alpha_s)$ is $\Gamma^{0}_{LL}(N,\alpha_s)$
minus its one--loop component, which appears in
$\Gamma^{0,l,+}(N)$.
$\alpha_s\Gamma^1_{2L}(N,\alpha_s)$ is the 
leading order in $\ln (1/x)$ term governing the evolution of $F_2(N,Q^2)$ in 
terms of $F_L(N,Q^2)$. It is a power of $\alpha_s$ up on 
$\Gamma^{0}_{LL}(N,\alpha_s)$, but is sufficiently leading (i.e. is not 
subleading to any other contributions) to make an 
appearance in the input terms in \fullsolfderiv\ and \fullsolfin. 
It is given by
\eqn\defgamtwol{\Gamma^1_{2L}(N,\alpha_s)={\gamma^1_{fg}(N,\alpha_s)+
\gamma^0_{gg}C^1_{2,g}(N,\alpha_s)\over 2\pi C^1_{L,g}(N,\alpha_s)},}
and thus is equal again to one of the physical anomalous dimensions in \Cat.  
($\Gamma^{1,0}_{2L}(N)$ is the one--loop contribution
to $\Gamma^1_{2L}(N,\alpha_s)$, and must be subtracted in some places in 
\fullsolfderiv\ and \fullsolfin\ in order to avoid double counting.) 
However, once again we stress that these anomalous dimensions are not needed 
to derive these expressions, but that they naturally appear in the 
end results. 

The term $\exp[\ln(Q_0^2/A_{LL})\Gamma^0_{LL}(N,\alpha_s(Q_0^2))]$ appears in 
some of the input parts in our expressions. This leading $\ln(1/x)$ 
contribution to the inputs is derived by demanding that the form of 
our expressions is invariant under changes in the arbitrary starting
scale $Q_0^2$ at the order at which they are calculated, i.e. the expression 
for the structure functions as a whole is genuinely of leading order.
It is easy to see that
the variation of this input term under a change in $Q_0^2$ cancels the
leading order change in the leading $\ln(1/x)$ evolution. 
This procedure leaves us 
with an unknown scale $A_{LL}$, but at this scale the inputs become 
the nonpertubative inputs, and we would therefore expect $A_{LL}$ to be 
the sort of scale where perturbation theory starts to break down. The other
terms in \fullsolfin\ proportional to $\ln(Q_0^2/A_{LL})$ are likewise 
required to make the full expression invariant under 
changes in $Q_0^2$ at leading order.

Having obtained the full leading order RSC
expressions for $(d\,F_2(N,Q^2)/d\,\ln Q^2)$ and $F_2(N,Q_0^2)$ we integrate
$(d\,F_{2}(N,Q^2)/d\,\ln Q^2)_{RSC,0}$ from $Q_0^2$ to $Q^2$ and add to 
the input $F_{2,RSC,0}(N,Q_0^2)$ in order to get our expression 
for $F_2(N,Q^2)$. This is essentially because it is the 
derivative of $F_2(N,Q^2)$ that begins at first order in $\alpha_s$ and
thus is a truly perturbative quantity. The difference between using this 
prescription and using the leading order $F_2(N,Q^2)$ directly is small. 

It is easy to check that under 
a change in the coupling, $\alpha_s\to \alpha_s+\delta\alpha^2_s$ the 
change of each of our expressions is of higher order in 
$\alpha_s$ than any terms appearing.
Thus, we have full leading order, including leading $\ln (1/x)$ terms, 
renormalization scheme consistent expressions for the structure functions. 
These are significantly different from both the one--loop expressions 
and the leading $\ln(1/x)$ expressions, although they clearly reduce to them 
in the appropriate limits. All terms in the expressions 
\fullsolfl-\fullsolfin\ are renormalization scheme as well as 
factorization scheme independent, as we would hope, and they
are the appropriate full expressions to use with the one--loop coupling 
constant.

Finally, we notice that this method of solution leads to a certain amount of
predictive power. We know the precise form of the 
structure function inputs in terms
of the nonperturbative inputs (which we imagine should be quite flat
at small $x$). Thus, up to the absolute normalization and the 
scale $A_{LL}$, we have predictions for the small $x$ form of the inputs for 
$F_L(N,Q^2)$ and $F_2(N,Q^2)$ (as well as for $(d\,F_2(N,Q^2)/d\ln Q^2)$).
The normalization is fairly well set by the large $x$ data, and
we would expect $A_{LL} \lsim 1\Gev^2$, so there is an estimate for
the small $x$ form of each input. Moreover, the unknown elements are the 
same for each input, so there is a strong prediction for the 
relationships between the small $x$ inputs. However, the scale $Q_0^2$ which 
should be chosen is not determined. Nevertheless, 
it is a considerable consistency 
requirement that the relationships should be true for any choice of 
$Q_0^2$, and 
hopefully they can be well satisfied over a wide range of $Q_0^2$. 
In order for this to be true, $\alpha_s(Q_0^2)$ cannot be too sensitive
to $Q_0^2$, so we would not expect $Q_0^2$ to be particularly low. Also, since 
we can largely choose our structure functions at $Q_0^2$, but then have 
no freedom in how we evolve up and down in $Q^2$, we would imagine that 
when performing a fit it is advantageous if $Q_0^2$ is near the centre 
of the range of our data. 

\newsec{Fits to The Data.}

We use the expressions \fullsolfl-\fullsolfin\ to calculate the $x$-space
singlet structure functions. The nonsinglet structure functions are 
calculated using the normal one--loop prescription. By combining the
singlet and nonsinglet components and varying the free parameters 
($Q_0^2$,  $A_{LL}$, and the soft 
inputs for $F_L(x,Q^2)$ and $F_2(x,Q^2)$), we obtain the  
best fit for the available $F_2$ structure function data.\foot{In practice 
a variant of the program used 
by MRS is used for the fit, and inputs for the gluon and quarks of the
standard form are specified. To calculate structure functions
we use an effective factorization scheme which is of DIS type for
$F_2(x,Q^2)$, where the longitudinal coefficient functions are the one--loop
expressions, and the resummed anomalous dimensions 
and small $x$ inputs are chosen to produce
expressions for the structure functions matching \fullsolfl-\fullsolfin\
as closely as possible.
For simple inputs where exact analytic expressions for the $x$-space forms
of \fullsolfl-\fullsolfin\ can be found checks are made with the results 
of the evolution program, and any discrepancies are always much smaller 
than the errors in the data.} We note that 
the input $F^S_2(x,Q_0^2)$ and the evolution $d\,F_2(x,Q^2)/d\ln Q^2$
are forced by \fullsolfderiv\ and \fullsolfin\ to be trivially related
at small $x$, which is not the case when working at fixed order in $\alpha_s$.
The one--loop value for $\Lambda^{N_f=4}$ is chosen to be $100 \hbox{\rm MeV}$.
This precise value is not determined by a best fit, but a value near
this is certainly favoured. The published values of 
$F_2(x,Q^2)$ are altered 
to take account of the fact that our predictions for $F_L(x,Q^2)$ 
are not the same as (i.e. are somewhat lower than) 
those used by H1 and ZEUS in their determination of $F_2(x,Q^2)$.
Thus, the $F_2(x,Q^2)$ values are a little lower for the largest values 
of $Q^2/x$ than in \hone\ and \zeus.  

We treat the heavy quark thresholds rather naively. Taking $m_c^2=3\Gev^2$ 
and $m_b^2=20\Gev^2$, we simply change the number of active quark 
flavours discontinuously at these values of $Q^2$: since $F^S_L(N)$
has a large component proportional to
$N_f$,  $F_L(x,Q^2)$ is discontinuous at these values of $Q^2$,
as is $d\,F_2(x,Q^2)/d\ln(Q^2)$. $\alpha_s(Q^2)$ is continuous at the 
thresholds, being defined by
\eqn\defcoup{\alpha_{s,n}(Q^2)=\alpha_{s,n+1}(Q^2)\biggl(1+
{\alpha_{s,n+1}(Q^2)\over 6\pi}\ln(m_{n+1}^2/Q^2)\biggr).}
This treatment of quark thresholds is consistent with the 
decoupling theorem, in so much that it guarantees the 
correct expressions far above or below thresholds \ref\dec{J.C. Collins
and W.K. Tung, \NP B \vyp{278}{1986}{934}.}. It is clearly 
unsatisfactory near the thresholds and must be improved. 
However, the prescription has little effect on $F_2(x,Q^2)$ 
in the region of the fit: in the curves for $F_2(x,Q^2)$ the $b$-quark 
threshold is barely noticeable, while the kink at the $c$-quark threshold 
only really affects a handful of data points at very small $x$, 
tending to hinder the fit (see \fig\data{The curves correspond to
the value of the proton structure function $F_2(x,Q^2)$ obtained from the 
leading order renormalization scheme consistent (LO(x)) calculation at 12
values of $x$ appropriate for the most recent HERA data. For clarity of 
display we add $0.5(12-i)$ to the value of $F_2(x,Q^2)$ each time the value 
of $x$ is decreased, where $i=1\to12$. The data are assigned to the $x$
value which is closest to the experimental $x$ bin (for more details see
the similar figure displaying the two--loop fits
in \MRSii). E665 data is also shown on the curves with 
the five largest $x$ values. The H1 and ZEUS data are normalized by $0.995$ 
and $1.01$ respectively in order to produce the best fit.}).   

The result of the best fit using the leading order, 
including leading $\ln (1/x)$ terms, RSC expressions 
(henceforth refered to as LO(x)) with 
$Q_0^2=40\Gev^2$ is compared with fits obtained using the 
standard two--loop method, where R$_1$ allows 
$\Lambda^{N_f=4}_{{\overline{\rm MS}}}$ to be free (giving 
$\Lambda^{N_f=4}_{{\overline{\rm MS}}}=241{\hbox{\rm MeV}}$) and 
R$_2$ fixes $\Lambda^{N_f=4}_{{\overline{\rm MS}}}=344{\hbox{\rm MeV}}$
to force a better fit to the HERA data. The new NMC data \ref\NMC{NMC 
Collaboration, hep-ph/9610231, to be published in \NP B.} for $F^{\mu p}_2$
and $F^{\mu d}_2$ is used with a lower $Q^2$ cut of $2\Gev^2$.\foot{The 
MRS fits are not performed
again: the $\chi^2$ for the new data is calculated using the same input
parameters in \MRSii\ (there is little indication that these would 
be changed much by the new data).} The results are shown in table 1
(full references for the experimental data can be found in \MRSii). As one can 
see, the LO(x) scheme independent fit is much better for the HERA data (even
when compared to R$_2$), much better for the BCDMS data (even when compared
to R$_1$) and similar in standard for the rest of the data. The overall fit is 
$\sim 200$ better for the whole data set. The results of the fit to the 
small $x$ data is shown in \data. 

The leading order renormalization scheme consistent 
expressions clearly provide a very good fit to the $F_2(x,Q^2)$ data.  
The fit shown is for the particular starting scale $Q_0^2=40\Gev^2$,
but the quality of the fit is extremely insensitive to changes in this scale
(where we allow $A_{LL}$ to be a free parameter for each $Q_0^2$), 
as we expect from the method of construction of the solutions. The fit is
essentially unchanged over the range $20-80 \Gev^2$, 
and we choose $40\Gev^2$ as the (logarithmically) central value. 
When $Q_0^2$ drops below $20\Gev^2$ the fit immediately gets markedly worse  
because of the discontinuity in $d\,F^S_2(x,Q^2)/d\ln Q^2$ at the threshold: 
$d\,F^S_2(x,Q^2)/d\ln Q^2$ suddenly becomes too large at small 
$x$ if $F^S_2(x,Q_0^2)$ is the correct size to fit the data.
The quality of the fit gets continuously
worse as $Q_0^2$ lowers further, becoming completely uncompetitive long
before reaching $m_c^2$.  We expect that a correct treatment of 
quark thresholds would lead to a smooth falling off of the quality of the 
fit, but that it would begin to deteriorate somewhere
in the region of $20\Gev^2$, due to $\alpha_s(Q_0^2)$ becoming too 
large below this value. 

The parameter $A_{LL}$, which should be a scale typical of soft 
physics, turns out to be $0.4\Gev^2$ for the fit starting at $Q_0^2=40
\Gev^2$. This decreases a little as $Q_0^2$ increases and {\it vice versa}.
For $Q_0^2=40\Gev^2$ the soft inputs for the fit are roughly
\eqn\softinputs{\hat F^S_L(x)\approx 2.65(1-x)^5, \hskip 1in F^S_2(x)\approx
(1-x)^{4}(1-0.6x^{0.5}+7x),}
where they have been forced to be flat as $x\to 0$. Allowing instead an 
asymptotic behaviour $x^{\lambda}$, where $\lambda \lsim 0.08$, leads to an
equally good fit. The importance of the leading $\ln(1/x)$ terms can be judged
by how they affect the fit. If, after obtaining the best fit, all terms
other than those in the one--loop expressions are set to zero, the quality 
of the fit is unchanged above $x=0.3$, begins to 
alter slightly below this, and is clearly much worse by the time we reach 
$x=0.1$. Thus, the leading $\ln(1/x)$ terms are important by this 
value of $x$.

It is not yet possible to extend the RSC
calculation beyond the leading order due to lack of knowledge of NLO in 
$\ln (1/x)$ terms. There is hope that these will shortly become available
\ref\NLOBFKL{L.N. Lipatov, and V.S. Fadin, \SJNP \vyp{50}{1989}{712}\semi
V.S. Fadin, R. Fiore and A. Quartarolo, \PR D \vyp{53}{1996}{2729}\semi
V. Del Duca, \PR D \vyp{54}{1996}{989}\semi
G. Camici and M. Ciafaloni, Preprint DFF-250-6-96.}, and 
when they do the NLO versions of \fullsolfl-\fullsolfin\ can be derived and
put to use. Only then should the NLO coupling constant be used in any fit.
As shown at leading order 
the $\ln(1/x)$ terms not present in the one--loop
expressions become important above $x=0.1$. However, much of this effect 
is due to the terms at ${\cal O}(\alpha_s^2)$, so the NLO
expression at fixed order in $\alpha_s$ should be a good approximation to
the full NLO RSC expression for $x$ somewhat 
lower than $0.1$, perhaps as low as $\sim 0.05-0.01$. 
However, until the full renormalization scheme consistent 
NLO expressions become available, we believe that it is premature to use 
fits to small $x$ structure function data to determine the NLO coupling 
constant (unless, of course, direct measurements of $F_L(x,Q^2)$ and other
less inclusive quantities at very small $x$ turn out to 
verify standard two-loop predictions).
The fixed order in $\alpha_s$ expressions should be accurate 
for CCFR, BCDMS and NMC data, which after all are still much more 
precise than HERA data, and fits to these data alone will provide the 
best determination of the NLO $\alpha_s(Q^2)$.

\newsec{Conclusion}

In this paper we have demonstrated that it is possible to
derive expressions for the structure functions which incorporate the 
leading $\ln(1/x)$  terms in a way which is renormalization scheme consistent, 
and as a consequence avoids any factorization scheme dependence. 
We have also shown that these full leading order RSC 
expressions lead to very good fits to the data and that, futhermore, they 
are able to do so using as inputs only soft distributions
for the singlet quark and
gluon, i.e. all powerlike behaviour is generated perturbatively, and 
determined in terms of the nonperturbative flat inputs and a 
soft scale $A_{LL}$.
Hence, this approach provides an explanation for the form of
the small $x$ structure function rather than just a way of fitting it.
Futher details of both the theory and fits, as well as other related issues,
are presented in \rst. It is certainly true that
the calculations must be improved to take account of massive
quark thresholds in a better manner, and work towards this end is in progress.
Nevertheless, with the present treatment we feel
that the quality of the fit and the degree of explanatory (if not predictive)
power, not to mention the scheme independence, give strong
justification for using this approach.

However, the quality of the fit alone is certainly not such a substantial
improvement on  more
standard approaches that it necessarily 
convinces one that this approach must be correct. In 
order to obtain some degree of verification we must obtain more 
experimental data. So far we have only probed $F_L(x,Q^2)$ indirectly, 
i.e. it is simply related to the derivative of $F_2(x,Q^2)$
(as well as to the input $F_2(x,Q_0^2)$ using this method). 
Having tied down the nonperturbative inputs and $A_{LL}$ and $Q_0^2$ from
our fit to $F_2(x,Q^2)$, we have a prediction for $F_L(x,Q^2)$. The result of
this prediction for the fit with $Q_0^2=40\Gev^2$
is shown in \fig\fl{Comparison of predictions for $F_L(x,Q^2)$ using the full
renormalization scheme consistent (LO(x)) fit and the two--loop MRSR$_1$ 
fit. For both sets of curves $F_L(x,Q^2)$ increases with increasing $Q^2$ 
at the lowest $x$ values.}, where it is compared to the prediction 
using the NLO in $\alpha_s$ approach and the 
MRSR$_1$ fit. As one can see, it is smaller
than the MRSR$_1$ $F_L(x,Q^2)$, but becomes steeper at very small $x$. The
prediction for $F_L(x,Q^2)$ is weakly dependent on the value of $Q_0^2$ 
chosen: the value at $Q^2=5\Gev^2$ and $x=10^{-4}$
varies by $\pm 10\%$ within our range of 
$Q_0^2$ (increasing with $Q_0^2$), and by
less than this for higher $x$ and $Q^2$.\foot{The author has submitted 
two conference proceedings on the topic of the current 
paper \ref\rstproc{R.S. Thorne, proceedings of DIS 96, (Rome, April 1996); 
proceedings of workshop ``Future physics at HERA'', (DESY, Hamburg, 1996).}  
and should point out that these are both incomplete. In the former 
the expression 
\fullsolfin\ was not used for the input for $F_2(x,Q^2)$. Thus, the fit 
imposed far less constraint on $Q_0^2$ than
the full procedure, and a value of $5\Gev^2$ was used, which resulted in a
prediction of $F_L(x,Q^2)$ that is much too small. The latter claimed that an 
input for $F_L(x,Q^2)$ a little smaller than that consistent with the full
set of expressions was needed for the best fit,
even at the optimum $Q_0^2$. This was due to 
there being no account whatever taken of the $b$-quark threshold. The
correction makes very little difference to phenomenological
results.} The very recent results on $F_L(x,Q^2)$ 
for $0.01\gsim x \gsim 0.1$ from NMC \NMC\ are matched 
far better by the LO(x) $F_L$ than the MRSR$_1$ $F_L$ (the latter being 
rather large). However, it is fair to say that any problems with the 
MRSR$_1$ $F_L$ can very probably be assigned to the treatment of the charm 
quark threshold, i.e. the predicted $F_L(x,Q^2)$ in the last of \GRV\ 
matches the data well. Measurements of $F_L(x,Q^2)$ at $x<10^{-2}$
would be a better discriminant between fixed order in $\alpha_s$ calculations
and those involving leading $\ln (1/x)$ terms. 
However, the sort of ``determination'' of $F_L(x,Q^2)$
already performed by H1 \ref\fldet{H1 collaboration, paper pa02-069, 
submitted to ICHEP 1996, (Warsaw, July, 1996).} is really 
only a consistency check for a particular 
fit, and is by no means a true measurement of $F_L(x,Q^2)$.
Real, direct measurements of $F_L(x,Q^2)$ at HERA would be an important 
(and probably essential) way of determining the validity of the 
approach in this paper, and the genuine importance of leading 
$\ln(1/x)$ terms in structure functions.

\medskip
\noindent{\bf Acknowledgements.}
\medskip
I would like to thank R.G. Roberts for continual help during the period of 
this work and for the use of the MRS fit program. I would also like to thank 
Werner Vogelsang and Mandy Cooper--Sarkar for helpful discussions.

\vfill 
\eject

\noindent Table 1\hfil\break
\noindent Comparison of quality of fits using full leading order (including 
$\ln (1/x)$ terms) renormalization scheme consistent expression, LO(x), and
two--loop fits MRSR$_1$ and MRSR$_2$. For the LO(x) fit the H1 data is 
normalized by a factor of $0.995$, the ZEUS data by $1.01$, the BCDMS data
by $0.98$, the CCFR data by $0.95$, and the rest by $1.00$.  
\medskip

\hfil\vtop{{\offinterlineskip
\halign{ \strut\tabskip=0.6pc
\vrule#&  #\hfil&  \vrule#&  \hfil#& \vrule#& \hfil#& \vrule#& \hfil#&
\vrule#& \hfil#& \vrule#\tabskip=0pt\cr
\noalign{\hrule}
& Experiment && data && \omit &\omit&$\chi^2$&\omit& \omit &\cr
&\omit&& points && LO(x) &\omit& R$_1$ &\omit& R$_2$&\cr
\noalign{\hrule}
& H1 $F^{ep}_2$ && 193 && 128 && 158 && 149 &\cr
& ZEUS $F^{ep}_2$ && 204 && 256 && 326 && 308 &\cr
\noalign{\hrule}
& BCDMS $F^{\mu p}_2$ && 174 && 190 && 265 && 320 &\cr
& NMC $F^{\mu p}_2$ && 129 && 124 && 163 && 135 &\cr
& NMC $F^{\mu d}_2$ && 129 && 109 && 134 && 99 &\cr
& NMC $F^{\mu n}_2/F^{\mu p}_2$ && 85 && 142 && 136 && 132 &\cr
& E665 $F^{\mu p}_2$ && 53 && 8 && 8 && 8 &\cr
\noalign{\hrule}
& CCFR $F^{\nu N}_2$ && 66 && 51 && 41 && 56 &\cr
& CCFR $F^{\nu N}_2$ && 66 && 49 && 51 && 47 &\cr
\noalign{\hrule}}}}\hfil

\footatend\vfill\supereject\immediate\closeout\rfile\writestoppt
\baselineskip=14pt\centerline{{\bf References}}\bigskip{\frenchspacing%
\parindent=20pt\escapechar=` \input refs.tmp\vfill\eject}\nonfrenchspacing

\vfill\eject\immediate\closeout\ffile{\parindent40pt
\baselineskip14pt\centerline{{\bf Figure Captions}}\nobreak\medskip
\escapechar=` \input figs.tmp\vfill\eject}

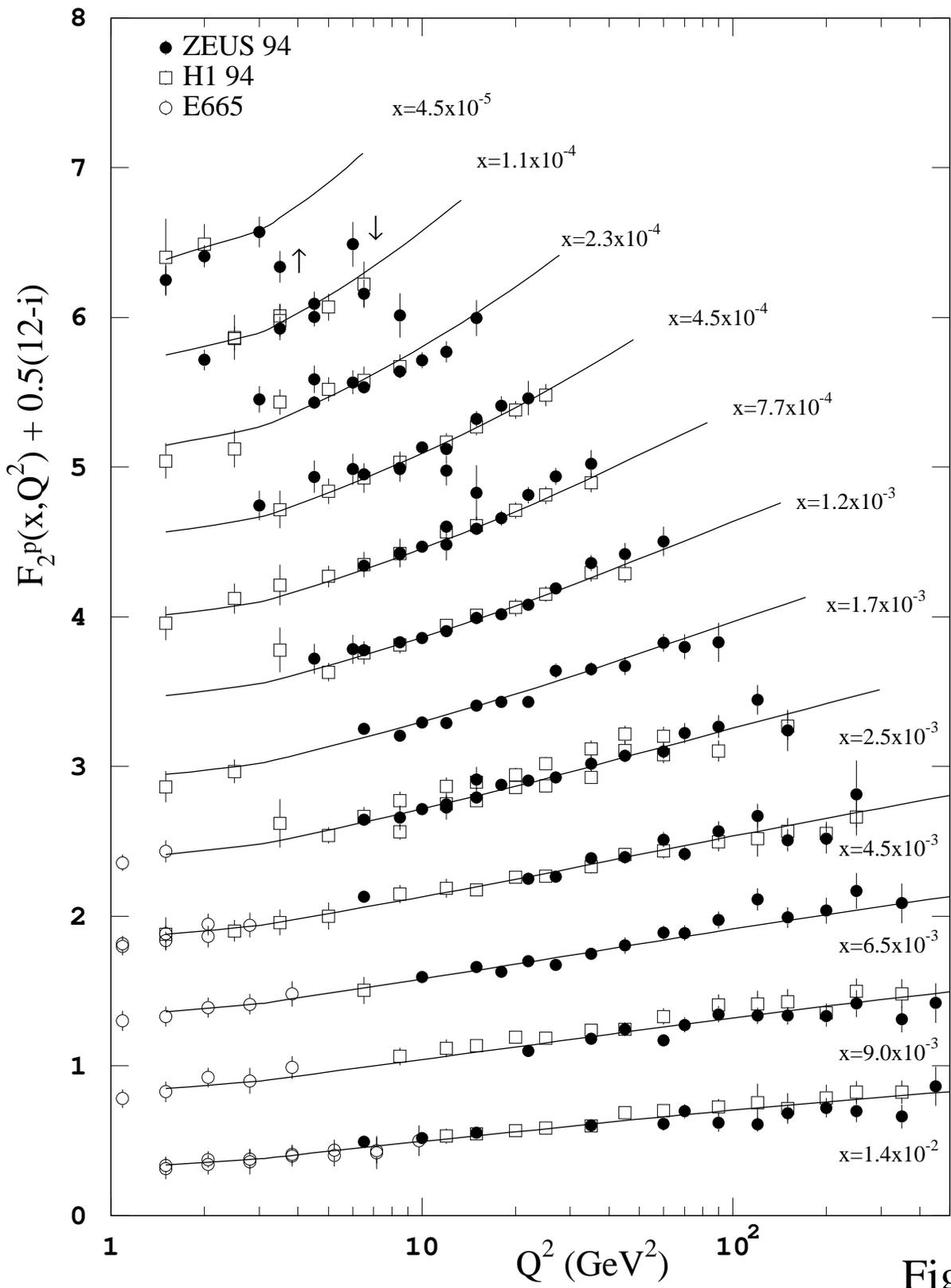
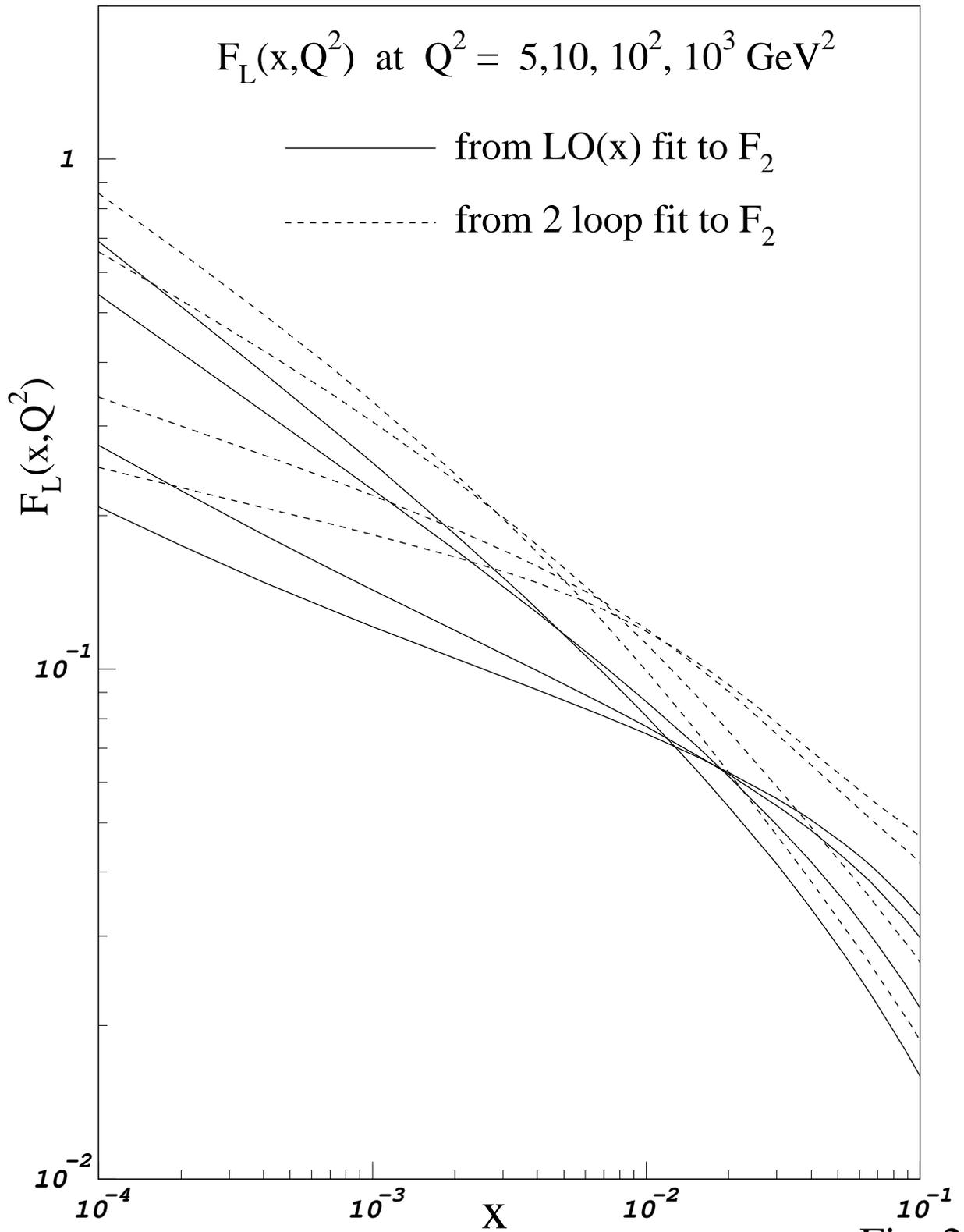
\end